\documentclass[10pt]{article}
\usepackage{amsmath}
\usepackage{amssymb}
\usepackage{amscd}
\usepackage{theorem}
\usepackage{xspace}
\input xy
\xyoption{all}

\def\cala{{\cal A}}

\def\calh{{\cal H}}
\def\call{{\cal L}}

\def\demi{{\frac{1}{2}}}

\def \rvac {\!\! \left. \left. \right| \! 0 \right\rangle}

\def\pb{\mathfrak {pB}}
\def\pf{\mathfrak {pF}}
\def\pbq{{\mathfrak {pB}}_{q}}
\def\pfq{{\mathfrak {pF}}_{q}}

\newtheorem{theorem}{THEOREM}
\newtheorem{lemma}{LEMMA}
\newtheorem{proposition}{PROPOSITION}

\newtheorem{definition}{DEFINITION}
\newcommand{\beq}{\begin{equation}}
\newcommand{\eeq}{\end{equation}}
\newcommand{\ba}{\begin{array}}
\newcommand{\ea}{\end{array}}
\newcommand{\beqa}{\begin{eqnarray}}
\newcommand{\eeqa}{\end{eqnarray}}

\begin{document}
\vspace*{0.3cm}
                \begin{center}
                { \LARGE Hopf Structure and Green Ansatz
     of  \\[6pt] Deformed Parastatistics Algebras}  \\[15pt]
\vspace{0.3cm}
 { 
 Boyka Aneva${}^{1,2}$   \hspace{40pt}       Todor  Popov${}^{1,3}$}
                \end{center}
\begin{center}
{ ${}^{1}$\small Institute for Nuclear Research and Nuclear Energy, \\
        \small  Bulgarian Academy of Sciences   \\
          \small bld. Tsarigradsko chauss\'ee 72,
                 \small BG-1784 Sofia,  Bulgaria }
\end{center}
\begin{center}
{\small ${}^{2}$ 
 Fakult\"at  f\"ur Physik,
Ludwig-Maximilians-Universit\"at M\"unchen,
\\
D-80333 M\"unchen, Germany }
\end{center}
\begin{center}
{\small ${}^{3}$ Laboratoire d'Informatique de l'Universit\'e
Paris-Nord, \\
        \small  CNRS (UMR 7030) and Universit\'e Paris-Nord  \\
 99 av. J-B Cl\'ement,  F-93430 Villetaneuse,  France }
\end{center}

\vspace{0.3cm}

    \begin{abstract}
        Deformed parabose and parafermi algebras
    are revised and endowed with  Hopf structure in a natural way.
    The noncocommutative coproduct allows for construction of
     parastatistics Fock-like representations, built out
    of the simplest deformed bose and fermi representations.
    The construction gives rise to quadratic algebras of deformed
    anomalous commutation relations which define the generalized
    Green ansatz.
    \end{abstract}
\vspace{0.5 cm}
    \section{Introduction}
         Wigner was the first to remark that the
canonical quantization was not the most general quantization scheme
consistent with the Heisenberg equations of motions \cite{W}.
Parastatistics  was introduced by Green \cite{Green} 
as a general
quantization method of quantum field theory different from the
cannonical  Bose and Fermi quantization. This generalized statistics
is based on two types of algebras with trilinear exchange relations,
namely the parafermi and parabose algebras.

The representations of the parafermi and parabose algebras are
labelled by a
non-negative integer $p$ - the order of parastatistics. The simplest
non-trivial representations arise for $p=1$ and coincide with the
usual Bose(Fermi) Fock
 representations. The states in a Bose(Fermi) Fock space are totally
symmetric(antisymmetric), i.e., they transform according to the one
dimensional
representions of the  {\it symmetric group}.
 Fock-like representations of parastatistics of order  $p\geq 2$
 correspond to higher-dimensional representations of the symmetric
group
 in the Hilbert space of multicomponent fields.


At the core of the 
interest in generalized statistics is (twodimensional) statistical
mechanics of phenomena such as fractional Hall effect, high-$T_c$
superconductivity. The experiments on quantum Hall
effect confirm the existence of fractionally charged excitations
\cite{QHE}.
Models with fractional statistics and infinite statistics have been
explored,
termed as anyon statistics \cite{anyons} and quon statistics
\cite{quons}.

The attempts to develop
nonstandard quantum statistics evolved naturally to the
study of deformed parastatistics algebras.
The guiding principle in these developments is the isomorphism between
the  
parabose algebra $\pb(n)$,
 parafermi algebra  $\pf(n)$ (with $n$ degrees of freedom)
  and the universal enveloping algebra of the orthosymplectic
algebra   $osp(1|2n)$, resp. orthogonal algebra $so(2n+1)$.
        The quantum counteparts $\pbq(n)$ and $\pfq(n)$ were defined
to be
        isomorphic {\it as algebras} to the { quantized universal
enveloping algebras} (QUEA)  $U_q(osp(1|2n))$\cite{Hadji},
        resp.   $U_q(so(2n+1))$\cite{Pal2}.

        In the present work we write a complete basis of defining relations of
the algebras   $\pbq(n)$ and $\pfq(n)$(see Theorem \ref{th1})
        extending what has been done in \cite{Hadji,Pal2,Pal1}.
The novelty with respect to the known definition
of deformed parastatistics is the system of  homogeneous relations (\ref{II}, \ref{II*}).
They allow to
continue the  isomorphism of the algebras as Hopf algebra
morphism  (see Theorem \ref{Hopfpf})
        which endows the deformed parastatistics algebra at hand with natural
Hopf structure. With the defined Hopf structure
the  parastatistics algebras
$\pbq(n)$ and $\pfq(n)$
  become isomorphic  { \it as   Hopf algebras} to the QUEA
$U_q(osp(1|2n))$ and
           $U_q(so(2n+1))$, respectively.

        The Green ansatz is intimately
related to the coproduct on the parastatistics algebras;
 it was realized that every  parastatistics algebra  representation
of arbitrary order $p$  arises through the iterated coproduct
\cite{Pal3}(see also \cite{Quesne}).
We make use of the noncocommutative coproduct on the Hopf
parastatistics algebras $\pbq(n)$ and $\pfq(n)$   to
         construct a quadratic algebra which is a deformation of
         the Green ansatz for the classical algebras    $\pb(n)$ and
$\pf(n)$.  \\

The paper is organized as follows.  
In section 2 we define
 the  relations of the quantized
parastatistics. 
Section 3 is devoted to the analysis of the Hopf algebra structure of
the proposed quantized parastatistics algebras.  In Section 4 we show
that the $q$-deformed bosonic (fermionic) oscillator algebra arises as
 the simplest non-trivial representation of the deformed
parastatistics.
 Further
in Section 5 
 the Green ansatz is generalized for the deformed parastatistics
algebras
  $\pbq(n)$ and $\pfq(n)$.

Throughout the text by an associative algebra we  mean an
associative algebra with unit $1$ over the complex numbers $\mathbb
C$.

     \section{Deformed Parastatistics Algebras}

We first recall the definitions of the parastatistics
algebras introduced by Green \cite{Green} as a generalization
of the  Bose-Fermi alternative.
\begin{definition} The  parafermi algebra $\pf(n)$
     (parabose algebra $\pb(n)$) is an
associative (super)algebra  generated by the creation $a^{+}_i$ and
annihilation
$a^-_i$ operators for $i=1,\ldots,n$   subject to  the relations
\beqa
  \ba{rcrcrcc}
  [\![ [\![a_{i}^{+},a_{j}^{-} ]\!], a_{k}^{+}]\!] &=&  2 \delta_{jk}
  a_{i}^{+}
  &\quad & [\![ [ \![a_{i}^{+}, a_{j}^{+} ]\!],a_{k}^{+} ]\!] &=&0
  \\[4pt]
  [\![[\![ a^{+}_i,a_{j}^{-} ]\!], a_{k}^{-}]\!] &=& - 2 \delta_{ik}
  a_{j}^{-}&
  \quad &[\![ [ \![a_{i}^{-}, a_{j}^{-} ]\!],a_{k}^{-} ]\!] &=&0
  \ea
  \label{1}
  \eeqa
 where $[\![a,b]\!]=ab - (-1)^{deg(a)deg(b)}ba$ is the supercommutator
 and all the generators of $\pf(n)$($\pb(n)$) are taken to be even $deg(a_{j}^{\pm})=\bar{0}$
 (odd $deg(a_{j}^{\pm})=\bar{1}$). ${}^{(}$\footnote
 {In this definition  only the linearly independent relations are written, other relations
follow from the (super-)Jacobi identities}
\end{definition}

The parafermi algebra 
is isomorphic to the universal
enveloping algebra
$U(so(2n+1))$
of the  orthogonal algebra $so(2n+1)$, $\pf(n) \simeq
U(so(2n+1))$\cite{KT} 
while the parabose algebra 
is isomorphic
to the universal enveloping algebra $U(osp(1|2n))$ of the
orthosymplectic superalgebra $osp(1|2n)$, $\pb(n) \simeq U(osp(1|2n))$\cite{GP}.

The idea of quantization of the parastatistics algebras
is to ``quantize'' the classical isomorphisms 
i.e.,
to deform the trilinear relations
(\ref{1}) in such a way that the arising deformed parafermi $\pfq(n)$
and parabose $\pbq(n)$ algebras are isomorphic to the
{ \it quantized universal enveloping  algebra} (QUEA)
of a Lie  (super)algebra \cite{Drin, Jimbo, FRT, KhTo} 
\beq
\pfq(n) \simeq U_{q}(so(2n+1)) \qquad \pbq \simeq U_{q}(osp(1|2n)).
\label{isoq}
\eeq

 The  proofs of the algebra isomorphisms  $\pbq \simeq
U_{q}(osp(1|2n))$\cite{Hadji} and
  $\pfq(n) \simeq U_{q}(so(2n+1))$\cite{Pal2}
has shown the equivalence of the  paraoscillator definition of
the $U_{q}(osp(1|2n))$ and $U_{q}(so(2n+1))$
with their definition in terms of  Chevalley generators.
In this way a minimal set of  relations
(a counterpart of the Chevalley-Serre relations)
is  obtained  providing an algebraic  (but not linear) basis
of the defining ideal of the QUEA at hand.

  We are interested in a complete description  of  the defining
	 ideal  for the parastatistics algebras (i.e., the counterpart of the Cartan-Weyl  definition of the QUEA).
This is not only  a question of  pure academic interest, our motivation
came from the study of the Hopf algebraic structure on the
parastatistics algebras which to the best of our knowledge
was studied only for some particular cases
(see \cite{Pal4} for $\pbq(2)$). The complete basis of relations
is generated from the known algebraic one and allows for endowing the $\pfq(n)$ and $\pbq(n)$
with a Hopf algebra structure.
We now sketch the procedure
of deriving the complete $U_{q}$-linear basis for the parastatistics algebras.

The Lie  superalgebra $osp(1|2n)$,
denoted  as $B(0|n)$ in the Kac table \cite{Kac}
has the same Cartan matrix 
as
the simple  $B_n$ algebra $so(2n+1)$.
The  Chevalley-Serre  relations of QUEAs $U_{q}(so(2n+1))$ and $U_{q}(osp(1|2n))$
with generators  $q^{\pm H_{i}}\equiv q^{\pm H_{\alpha_{i}}}$ and
$E_{\pm i}\equiv E^{\pm \alpha_{i}}$, corresponding to the simple roots $\alpha_i$, read
\beq
\ba{lll}
q^{H_{i}}q^{\pm H_{j}}=q^{\pm H_{j}}q^{H_{i}} &
q^{H_{i}} E_{\pm j} q^{-H_{i}}= q^{\pm a_{ij}} E_{\pm j } &   1\leq i,j
\leq
n \\[6pt]
[2] [E_i,E_{-j}]=\delta_{ij} [2H_{i}] & [\![E_n,E_{-n}]\!] = [2H_{n}]&  1\leq i\leq
n-1 \\[6pt]
[E_{\pm i}, E_{\pm j}] = 0   &&     |i-j|\geq 2
\\[6pt]
[E_{\pm (i+1)}, [E_{\pm (i+1)}, E_{\pm i} ]_{q} ]_{q^{-1}} =0 &
 &
 1\leq i\leq n-2\\[6pt]
 [E_{\pm i}, [E_{\pm i}, E_{\pm (i+1)} ]_{q} ]_{q^{-1}} =0 \, \, \, =&
 [ [\![ [ E_{\pm (n-1)},E_{\pm n} ]_{q^{-1}},E_{\pm n} ]\!], E_{\pm
n}]_{q}
 &
 1\leq i\leq n-1
\ea
\label{quea}
\eeq
where $[x,y]_q=xy-qyx$ is the $q$-commutator,
$\alpha_n$ is the only odd simple root of $osp(1|2n)$
and $a_{ij}=(\alpha_i,\alpha_j)$ is the symmetrized Cartan matrix (same for both cases) given by
$a_{ij}=2\delta_{i j}-\delta_{i n}-\delta_{i+1j}-\delta_{i j+1}$.
The quantum bracket is
chosen to be
$[x] := \frac{q^{\frac{x}{2}}- q^{-\frac{x}{2}} }{q^{\frac{1}{2}}-
q^{-\frac{1}{2}}}$. 

The essential point in the proof of the isomorphism is the change of
basis for the QUEA by choosing the orthogonal system of
roots $\varepsilon_{i}$
as an alternative of the simple roots.
The ladder operators $E^{+ \varepsilon_{i}}$ and $E^{- \varepsilon_{i}}$
related to the roots $\varepsilon_{i}$  are the parastatistics creation
and annihilation operators $a^{+}_i$ and  $a^{-}_{j}$
\cite{Hadji,Pal1}
and the  
change
$\varepsilon_{i}=\mathop{\sum}_{k=i}^{n} \alpha_{k}$
implies
\beq
\ba{rcl}
a^{+}_i&=& [E_{i},[E_{i+1}, \ldots [E_{n-1},E_{n}]_{q^{-1}} \ldots
]_{q^{-1}} ]_{q^{-1}} \\
a^{-}_{i} &=& [[\ldots[E_{-n},E_{-n+1}]_{q} \ldots,E_{-(i+1)}]_{q},
E_{-i} ]_{q}
  \ea
  \label{algiso}
\eeq
With the help of the inverse change
$\alpha_{i}=\varepsilon_{i} -  \varepsilon_{i+1}$, $i < n$, and
$\alpha_{n}= \varepsilon_{n}$
the corresponding change of basis on the Cartan subalgebra reads
$H_{i}= h_{i}-h_{i+1}$, $i<n$, $H_{n}=h_{n}$.
By construction $q^{h_{i}}q^{h_{j}}=q^{h_{j}}q^{h_{i}}$.
The inverse change of basis allows to express the Chevalley ladder
generators as
\beq
\ba{rclcrcl}
E_{i} &=& \frac{1}{[2]} q^{-h_{i+1}}[\![a^{+}_i, a^{-}_{i+1}]\!]  &&
E_{-i} &=&  \frac{1}{[2]} [\![ a^{+}_{i+1}, a^{-}_{i}]\!] q^{h_{i+1}}\qquad
  i < n  \\ [8pt]
  E_{n}&=&  a^{+}_{n}  && E_{-n}&=&  a^{-}_{n}
\ea
\label{inviso}
\eeq

 The complete basis (over $\mathbb C(q)$) of relations defining
 the deformed parastatistics algebras, i.e., the analog of (\ref{1})
 is given by  the following
\begin{theorem}
   The  deformed parafermionic $\pfq(n)$ (parabosonic  $\pbq(n)$)
  algebra is  the associative (super)algebra  generated by the
  creation and annihilation operators  $a^{\pm}_{i}$ and  Cartan generators $q^{\pm h_{i}}$ for $i=1,\ldots ,n$
 subject to the relations 
 \beqa
  q^{h_{i}}a^{\pm}_{j}q^{-h_{i}} = q^{\pm \delta_{ij}} a^{\pm }_{j} &&  \quad
	[\![a^{+}_{i},a^{-}_{i} ]\!]= \frac{q^{h_{i}}- q^{-h_{i}} }{q^{\frac{1}{2}}-
q^{-\frac{1}{2}}} = [2h_{i}]
\label{fe}
\\[4pt]
[\![ [\![a^{+}_{i},a_{j}^{-} ]\!], a^{+}_{k}]\!]_{q^{-\delta_{ik}\sigma_{j,k}}}
&=& \quad [2] \hspace{3pt} \delta_{jk} a^{+}_{i}
q^{\sigma_{i,j}h_{j} }
+ (q-q^{-1})\theta_{i,j;k} a^{+}_{i}[\![a^{+}_{k},a_{j}^{-} ]\!]
,\, \label{I} \\[4pt]
[\![[\![a^{+}_{i},a_{j}^{-} ]\!], a^{-}_{k}]\!]_{q^{-\delta_{jk}\sigma_{i,k}}}
&=& - \hspace{3pt} [2]\delta_{i k} a^{-}_{j}
q^{-\sigma_{i,j}h_{i} } -
(q-q^{-1})\theta_{j,i;k} [\![a^{+}_{i},a_{k}^{-}]\!]  a^{-}_{j}
\label{I*}
\eeqa
   together with the analogues of the Serre relations
   \beqa
   [\![  [\![ a^{\pm}_{ i_1},a^{\pm}_{ i_3} ]\!], a^{\pm}_{ i_2}
]\!]_{q^{2}} +
   q[\![ [\![a_{i_1}^{\pm}, a_{i_2}^{\pm}   ]\!], a_{i_3}^{\pm}]\!]
	    =0&\qquad & i_1<i_2\leq i_3 \label{II}
    \qquad   \\ [4pt]    [\![ a_{i_2}^{\pm} ,
[\![ a_{i_1}^{\pm},a_{i_3}^{\pm} ]\!]]\!]_{{q^2}}
   +{q} [\![  a_{i_1}^{\pm}, [\![ a_{i_2}^{\pm},  a_{i_3}^{\pm}]\!]]\!]
	    =0&\qquad & i_1 \leq i_2<i_3 \label{II*}
   \eeqa
 where all the  generators $a^{\pm}_{i}$  are
 taken to be even,  $deg(a^{\pm}_{i})=\bar{0}$ (odd,
  $deg(a^{\pm}_{i})=\bar{1}$) and
the symbols $\theta_{i,j;k}$, $\sigma_{i,j}$ stay for
  $\theta_{i,j;k}=\frac{1}{2}\epsilon_{ij} \epsilon_{ijk} (\epsilon_{jk} -
   \epsilon_{ik})$,
   $\sigma_{i,j}= \epsilon_{ij} + \delta_{ij}$ or $\sigma_{i,j}= \epsilon_{ij}
   - \delta_{ij}$.
    ${}^{(}$\footnote{${}^{)}$ $\epsilon$ is the
   Levi-Civita symbol with $\epsilon_{ij}=1$ for $i<j$.  The tensor
   $\theta_{i,j;k} =-\theta_{j,i;k}$ is vanishing except for $i<k<j$ and $i>k>j$
	  when it takes values $+1$ and $-1$,
   respectively.}
\label{th1}
\end{theorem}

 The deformed parastatistics algebras admit an anti-involution $\ast$ 
		\beq
(a^{\pm}_i)^{\ast}=a^{\mp}_i \qquad 
(q^{\pm  h_i})^{\ast} = q^{\mp  h_i}  \qquad (q)^{\ast}=q^{-1}
\label{star}
\eeq
	induced by the 
 anti-involution   on the Chevalley basis
$(E_{\pm i})^{\ast}=E_{\mp i}$,  $H_{i}^{\ast}= H_{i}$.

To prove the theorem we  make use of the
$R$-matrix FRT-formalism
for QUEA  $U_{q}(g)$ of  a simple (super-)Lie algebra $g$
(see \cite{FRT},\cite{KhTo}), 	introducing the $L$-functionals for
	$U_{q}(g)$ in the form of   upper
(lower)-triangular matrices $L^{(+)}$ ($L^{(-)}$)
 \beq
R^{(+)}L^{(\pm)}_{1}L^{(\pm)}_{2}=L^{(\pm)}_{2}L^{(\pm)}_{1}R^{(+)}
\qquad
R^{(+)}L^{(+)}_{1}L^{(-)}_{2}=L^{(-)}_{2}L^{(+)}_{1}R^{(+)}
\label{frt2}
\eeq
where  $L^{(\pm)}_{1}= L^{(\pm)}
\otimes 1$, $L^{(\pm)}_{2}=1 \otimes L^{(\pm)}$
and $R^{(+)}=PRP$ is the corresponding  $R$-matrix for $U_{q}(g)$.

The $(n+1)\times(n+1)$ minor $L_{ij}^{ (+)}$,  $1\leq i,j \leq n+1$
of the $(2n+1)\times(2n+1)$   matrix
$L^{(+)}$ for the QUEA $U_{q}(so(2n+1))$ and $U_{q}(osp(1|2n))$
is very simple when expressed in
       terms of the generators $a^{\pm}_i$
\beqa
{\small
\left(L^{(+)}_{ij}\right)_{\small{ 1\leq i,j \leq n+1}}  =
\left(
     \ba{cccccc }
      q^{h_{1}} & \omega  [\![a^{+}_1, a^{-}_{2} ]\!] & \omega [\![a^{+}_1,
      a^{-}_{3} ]\!]   & \ldots &\omega [\![a^{+}_1, a^{-}_{n} ]\!] &
      c a^{+}_1 \\[5pt]
      0     & q^{h_{2}}  & \omega [\![a^{+}_2, a^{-}_{3} ]\!] & \ldots &
      \omega [\![a^{+}_2, a^{-}_{n} ]\!] &    c a^{+}_2  \\ [5pt]
           0    &   0    &   q^{h_{3}} & \ldots  & \omega[\![a^{+}_3,
           a^{-}_{n} ]\!] & c a^{+}_3
           \\
           \vdots & \vdots & \vdots & \ddots & \vdots & \vdots \\
        0& 0 &0  &\ldots & q^{h_{n}} &c a^{+}_n  \\
              0& 0 & 0&\ldots & 0& 1
     \ea
     \right) }
     \label{L+}
    \eeqa
  where $\omega= q^{\demi}-q^{-\demi}$. The coefficient $c=
  q^{-\demi}(q-q^{-1})$.
	One has $(L^{(+)}_{ij})^{\ast}= L^{(-)}_{ji}$.

The relations (\ref{fe}), (\ref{I}), (\ref{I*})  involving the entries
of the minors of $L^{(\pm)}_{ij}$ (\ref{L+}) for $ 1\leq i,j \leq n+1$
 follow directly from the
RLL-relations(\ref{frt2}) with the corresponding $R$-matrix
upon restricting  the indices from $1$ to $n+1$.
The restriction is possible due to the {\it ice condition}\cite{isa}.

         We label the  LHS (up to scalars in $\mathbb C(q)$) of the homogeneous
relations (\ref{II},\ref{II*}) by
\beq
\ba{rclcl}
         \Lambda^{i_1,i_3}_{i_2}  &=&
   [\![  [\![ a^{+}_{i_1},a^{+}_{i_3} ]\!], a^{+}_{i_2}
]\!]_{q^{2}} +
   q[\![ [\![a^{+}_{i_1}, a^{+}_{i_2}   ]\!], a^{+}_{i_3}]\!]     &\mbox{with}
& i_1<i_2<i_3
   \\ [4pt]
 \Lambda^{i_1,i_2}_{i_2}  &=&
ÿ         [\![[\![ a^{+}_{i_1}, a^{+}_{i_2} ]\!], a^{+}_{i_2}]\!]_{q}
&  \mbox{with} & i_1<i_2
 \\[4pt]
\tilde{\Lambda}^{i_1,i_2}_{i_3}  &=&
   [\![ a^{+}_{i_2} ,
[\![ a^{+}_{i_1},a^{+}_{i_3} ]\!]]\!]_{{q^2}}
   +{q} [\![  a^{+}_{i_1}, [\![ a^{+}_{i_2},  a^{+}_{i_3}]\!]]\!]
                & \mbox{with} & i_1<i_2<i_3   \\[4pt]
 \tilde{\Lambda}^{i_2,i_2}_{i_3}  &=&
  [\![  a^{+}_{i_2}, [\![ a^{+}_{i_2}, a^{+}_{i_3}
]\!]]\!]_{{q}}               & \mbox{with} & i_2<i_3
         \ea{}
         \label{hr}
                        \eeq

The QUEA $U_q(gl_n)$  has a natural inclusion in $U_q(so(2n+1))$ and
$U_q(osp(1|2n))$
being generated by  the Chevalley generators
$E_{\pm i}$, $1 \leq i \leq n-1$ and  $q^{\pm h_{i}}$.
(associated  with  the   $A_{n-1}$ subdiagram  in the $B_n$ Dynkin diagram).
 The inclusions $U_q(gl_n)\hookrightarrow \pfq(n)$ and $U_q(gl_n)
\hookrightarrow \pbq(n)$
 define an {\it adjoint} $U_q(gl_n)$-action  on $\pfq(n)$ and $\pbq(n)$
  (for $i \leq n-1$)
        $$
 ad_{E_i} a^{+}_j = [E_i, a^{+}_j]_{q^{\delta_{i j} -
\delta_{i+1\,j}}}=\delta_{i+1\,j} a^{+}_i  \qquad
   ad_{E_{-i}} a^{+}_j = [E_{-i}, a^{+}_j] q^{H_i} = \delta_{ij}
a^{+}_{i+1} \label{adj}
$$

 Let $\call$ denote   the space of
states   $\Lambda$ and $\tilde{\Lambda}$ where by states we mean  the cubic
polynomials
 determined from (\ref{hr}) up to
multiplication with scalars $\mathbb C(q)$.
The homogeneous relations (\ref{II},\ref{II*}) are $U_q(gl_n)$-covariant
 with respect to the
adjoint action. More precisely one has the following
\begin{lemma}The space $\call$ is   an irreducible finite-dimensional $U_q(gl_n)$-module
	 with lowest weight    $\Lambda^{n-1,n}_n$.(For the proof see the appendix.)

 \label{lamb}
 \end{lemma}
                                 The  distinguished state
$\Lambda^{n-1,n}_n$ expressed in terms of the Chevalley basis
 $E_i$
         is  the last Serre relation in (\ref{quea})
         $$
         \Lambda^{n-1,n}_n = [ [\![ [ E_{n-1},E_{ n}
]_{q^{-1}},E_{ n} ]\!], E_{ n}]_{q}=0
                 $$
         and thus $\Lambda^{n-1,n}_n$ has to be set to zero in
$\pfq(n)$  and $\pbq(n)$. Hence the whole representation $\call$ built
through the  $U_q(gl_n)$-adjoint action on
 the lowest weight  $\Lambda^{n-1,n}_n$ is trivial which
proves the homogeneous relations (\ref{II},\ref{II*}) for $a^+_i$.
The ones for $a^-_i$ follow by conjugation.

\section{Hopf structure on parastatistics algebras}

The QUE algebras  $U_{q}(so(2n+1))$ and $U_{q}(osp(1|2n))$
(\ref{quea})
endowed with the Drinfeld-Jimbo coalgebraic structure
\cite{Drin}, \cite{Jimbo}
\beq
\ba{lclclclr}
\Delta H_{i} &=& H_{i}\otimes  1 + 1 \otimes H_{i}  && S( H_{i}) &=
&-  H_{i}  & \epsilon(H_{i}) =0 \\
\Delta E_{ i} &=& E_{ i}\otimes 1  + q^{H_{i}}\otimes E_{ i}
  && S(E_{i})&=& - q^{- H_{i} } E_{i} & \epsilon(E_{ i})=0\\
\Delta E_{ -i} &=&  E_{ -i}\otimes q^{-H_{i}}  + 1\otimes E_{ -i}
  &&  S(E_{ -i})&=&- E_{ -i}q^{ H_{i} } &  \epsilon(E_{- i})=0
\label{drin}
\ea
\eeq
become  Hopf algebra and Hopf superalgebra, respectively.
${}^($\footnote {${}^)$For superalgebras $S$ is a graded antihomomorphism,
	$S(ab) = (-1)^{deg(a)deg(b)} S(b)S(a)$. }
One has  $S(x^{\ast})=S(x)^{\ast}$.

The isomorphism (\ref{isoq}) between the QUEA $U_q(so(2n+1))$
($U_q(osp(1|2n))$)
and the deformed parastatistics   algebra   induces a structure
of a Hopf (super)algebra on  $\pfq(n)$ ($\pbq(n)$).
One can formulate the following

\begin{theorem} The deformed parafermionic algebra $\pfq(n)$,
  the deformed parabosonic algebra $\pbq(n)$
         is  a Hopf algebra, a Hopf superalgebra, respectively when
   endowed with

          (i) a coproduct $\Delta$
      defined on the generators by
     $\Delta q^{\pm h_{i}}  = q^{\pm h_{i}}\otimes q^{\pm h_{i}}$
  \beqa
\Delta a^{+}_i &=& a^{+}_i \otimes 1 + q^{h_i} \otimes a^{+}_i +
\omega \mathop{\sum}\limits_{i< j\leq n}  [\![ a^{+}_i, a^{-}_j
]\!]
\otimes a^{+}_j \label{coa+}\\[4pt]
\Delta a^{-}_i &= & a^{-}_i \otimes q^{- h_i} + 1  \otimes a^{-}_i
- \omega  \mathop{\sum}\limits_{ i<j\leq n}
a^{-}_j  \otimes [\![ a^{+}_j, a^{-}_i]\!]
\label{coa-}
\eeqa

(ii) a counit $\epsilon$  defined on the generators by
 $\epsilon (q^{\pm  h_{i}})=1$,
$\epsilon(a^{\pm}_i)=0$

(iii) an  antipode $S$ 
defined on the
generators by
$S(q^{\pm h_{i}}) = q^{\mp h_{i}}$,
\beqa
 S(a^{+}_i) &=& - q^{-h_i}a^{+}_i -
\mathop{\sum}\limits_{s=1}^{n-i} (- \omega)^{s} \! \! \! \! \! \! \!
\! \! \!
\mathop{\sum}\limits_{i<j_1< \ldots <j_s\leq n}  \! \! \! \! \!
{W^{+}}^{i}_{j_1}{W^{+}}^{j_1}_{j_2} \ldots {W^{+}}^{j_{s-1}}_{j_s}
q^{-h_{j_s}}a^{+}_{j_s} \label{S+}\\
S(a^{-}_i) &=& \quad - a^{-}_i q^{h_i}-
\mathop{\sum}\limits_{s=1}^{n-i} (\omega)^{s} \! \! \! \!
\mathop{\sum}\limits_{ n \geq j_s > \ldots >j_1> i }
a^{-}_{j_s} q^{h_{j_s}}{W^-}^{j_s}_{j_{s-1}} \ldots
{W^-}^{j_2}_{j_1}{W^-}^{j_1}_{ i}
\label{S-}
\eeqa
where ${W^+}^{i}_{j}=  q^{-h_i} [\![ a^{+}_i,a^-_j ]\!]$,
${W^-}_{i}^{j}=   [\![ a^{+}_j,a^-_i ]\!] q^{ h_i}$ 
and $\omega=q^{\demi}- q^{-\demi}$.
\label{Hopfpf}
    \end{theorem}

  {\it \bf Proof:}
The Hopf structure on the elements of  $L^{(+)}$ and $L^{(-)}$
compatible with the Drinfeld
structure (\ref{drin}) (defined on the Chevalley basis) is given by
the coproduct $\Delta L^{\pm}$, the counit $\epsilon (L^{(\pm)})$ and the antipode  $S(L^{(\pm)})$
\cite{FRT}
\beqa
\Delta L_{ik}^{ (\pm)}= {\small \sum}_{j} L^{ (\pm)}_{ij}
\otimes
L^{( \pm)}_{jk} \qquad \epsilon (L_{ik}^{ (\pm)})= \delta_{ik} \qquad
{\sum}_{j} L_{ij}^{ (\pm)}S(L_{jk}^{ (\pm)}) =
\delta_{ik }^{} \label{coprL}
\eeqa

  (i)
  For the diagonal elements $L^{(+)}_{ii}  = q^{h_i}$
the coproduct formula in (\ref{coprL})  yields
\beq
\Delta(L^{(+)}_{ii}) =
\mathop{\sum}\limits_{1\leq j\leq 2n+1} L^{(+)}_{ij}
\otimes L^{(+)}_{ji} =
L^{ (+)}_{ii }\otimes L^{(+)}_{ii} =  q^{\pm h_{i}}\otimes q^{\pm
h_{i}}.
\eeq

  The coproduct of the elements  $L_{i\, n+1}^{(+)}$ when
$1\leq i \leq n$
  has the form
  \beq
\Delta L_{i\, n+1}^{(+)}= \mathop{\sum}\limits_{1\leq j\leq 2n+1}
L^{(+)}_{ij}
\otimes L^{(+)}_{j\,n+1}
=L_{i\, n+1}^{(+)}\otimes 1 +
\mathop{\sum}\limits_{i\leq j\leq n}  L_{ij}^{ (+)}\otimes
L_{j \, n+1}^{(+)}
\label{d}
\eeq
where we have used the triangularity of $L^{(+)}$ and
$L_{n+1\,n+1}^{(+)}=1$.
Inserting into eq.(\ref{d}) the values $L_{i\, n+1}^{(+)} =c a^{+}_i$
(\ref{L+})
and abridging the constant $c$ we get
\beq
\Delta a^{+}_i= a^{+ }_i\otimes 1 +
\mathop{\sum}\limits_{i\leq j\leq n} L_{ij}^{ (+)}\otimes a^{+ }_j
\eeq
which completes the proof of (\ref{coa+}) in view of (\ref{L+}).
Then  $\Delta a^{-}_i= (\Delta a^{+ }_i)^{\ast}$.

{\it (ii)} It follows from the definition of
the counit in (\ref{coprL}).

{ \it (iii)}
For the  diagonal elements the antipode formula in (\ref{coprL})
implies $S(L_{ii}^{(+)})=(L_{ii}^{(+)})^{-1}$ hence
$S(q^{\pm  h_{i}}) = q^{\mp h_{i}}$. 
For the nondiagonal elements  due to the triangularity of   $L^{(+)}$
 the antipode formula (\ref{coprL}) gives rise to the following
system of equations
\beq
\mathop{\sum}\limits_{i\leq j\leq n+1} L^{(+)}_{ij}
S(L^{(+)}_{j\,n+1})=\delta_{i\, n+1}
\quad \Longrightarrow \quad
\mathop{\sum}\limits_{i \leq j\leq n} L^{(+)}_{ij} S(L^{(+)}_{j\,n+1}) =
- L^{(+)}_{i\, n+1}.
\label{}
\eeq
Here we have made use of $S(L^{(+)}_{n+1\,n+1})= S(1) = 1$.
In view of $S(L^{(+)}_{i\,n+1})= c S(a^{+}_i)$ this is a linear
triangular system for $S(a^{+}_i)$ which
after normalisation takes the form
\beqa
S(a^{+}_i) + \omega
\mathop{\sum}\limits_{i< j\leq n} W^{i+}_{j} S(a^{+}_j) &=&-
q^{-h_i} a^{+}_i
\quad \mbox{where } \quad W^{i+}_{j} = q^{-h_i} [\![ a^{+}_i,a^-_j ]\!]
\eeqa
 and  the  solution of this system yields eq.(\ref{S+}).
The antipodes $S(a^{-}_{i})$  (\ref{S-}) are obtained through
  the conjugation,  $S(a^{-}_{i})=(S(a^{+}_i))^{\ast}$.
   $\Box$

This theorem is interesting in its own because it defines
the Hopf structure on another basis of generators for  QUEA
of the algebra $so(2n+1)$ and the superalgebra $osp(1|2n)$.

\section{ The oscillator representations}

The unitary representations $\pi_{p}$ of the parastatistics algebras
$\pb(n)$ and $\pf(n)$ (eq. \ref{1}) with  unique vacuum state
  are indexed by a non-negative integer $p$ \cite{GM}
(see also \cite{OK} and references therein).
The representation $\pi_{p}$ is the lowest weight representation with
a unique vacuum state $\rvac$  annihilated by all $a^{-}_{i}$
and labelled by the {\it order of  parastatistics } $p$
\beq
\pi_{p}(a^{-}_{i})\rvac = 0 \qquad  \qquad
\pi_{p}(a^{-}_{i})\pi_{p}(a^{+}_j)\rvac = p \delta_{ij}  \rvac, \quad \pi_{0}(x)\rvac =\epsilon(x)\rvac
\label{p}
\eeq
where the  {\it vacuum representation}, i.e., the trivial one corresponds to the counit $\epsilon$ of the Hopf parastatistics
algebra.
		 In the representation $\pi_{p}$ (\ref{p}) of the nondeformed
parastatistics algebras (\ref{1})
  the hamiltonian $\calh_i=h_{i}=\demi [a^{+}_{i},a^{-}_{i}]_{\mp}$ and the
number operator
  $N_{i}=a^{+}_{i}a^{-}_{i}$ associated to the
$i$-th paraoscillator
  are related by
\beq
\calh_i=h_{i}= N_{i} \mp \frac{p}{2}
\label{p2}
\eeq
where the upper (lower) sign is for parafermions (parabosons).

		  In the representation $\pi_{p}$ of the deformed parastatistics
algebras
the quantum analogue of the relation (\ref{p2}) holds
\[
[a^{+}_i,a^{-}_{i}]_{\mp}=[2]\calh_{i}
= [2h_{i}]=[2N_{i}\mp p]
\]
which implies  the deformed analogue of the $\pi_{p}$ defining
condition (\ref{p})
\beq
\pi_p(a^{-}_{i})\pi_p(a^{+}_j)\rvac = [p] \delta_{ij}  \rvac, \qquad
\pi_{0}(x)\rvac =\epsilon(x)\rvac
\label{pq}
\eeq
  The constant $\mp{[p]}/{[2]}$ plays the role
  of energy of the vacuum as
the constant $\mp{p}/{2}$ in (\ref{p2}) for the nondeformed algebras.

 The algebra of the
$q$-deformed fermionic (bosonic) oscillators $\mathfrak{F}_{q}(n)$
($\mathfrak{B}_{q}(n)$) arises as
  a  representation $\pi$
  of  order $p=1$ of the $\pfq(n)$($\pb_q(n)$) 
\beq
\left.
\begin{array}{rclrclc}
	{\underline a}^{+}_i{\underline a}^{+}_j \pm
q^{ \mp \epsilon_{ij}} {\underline a}^{+}_j {\underline a}^{+}_i &=&0    &
 {\underline a}^{-}_i {\underline a}^{-}_j \pm
q^{\mp \epsilon_{ij}} {\underline a}^{-}_j {\underline a}^{-}_i  &=&0   &\\[4pt]
{\underline a}^-_i {\underline a}^{+}_i \pm q  {\underline a}^{+}_j
{\underline a}^-_i
&=&q^{\pm {\underline N}_i} &
{\underline a}^-_i {\underline a}^{+}_i \pm q^{-1}  {\underline a}^{+}_i
{\underline a}^-_i
&=&q^{{\mp \underline N}_i}&\\[4pt]
		{\underline a}^{+}_i{\underline a}^{-}_j \pm
q^{ \mp \epsilon_{ji}} {\underline a}^{-}_j {\underline a}^{+}_i &=&0
		&	{\underline a}^{-}_i {\underline a}^{+}_j \pm
q^{\mp \epsilon_{ji}} {\underline a}^{+}_j {\underline a}^{-}_i    &=&0  &   i\neq j \\[4pt]
\end{array}  \right\}
\label{Bq}
\eeq
We have adopted the notaion
$\pi(x)=\underline{x}$ and use ${\underline N}_i={\underline h}_i \mp
\demi$.

The analysis \cite{unita} of the positivity of the norm for the
$\pbq(n)$ and $\pfq(n)$ representations in the simplest case $p=1$
shows
that such unitary representations
(realized as finite dimensional factor representions)
exist only for $q$ being a root of unity.

{\it Remark.} Unlike the case of
$\pbq(n)$ and $\pfq(n)$, the deformed relations of bosonic and fermionic
oscillator algebras (\ref{Bq}) do not define Hopf ideals.

\section{Green Ansatz}
The Green ansantz  was introduced by Green
in the same paper \cite{Green} in which he defined parastatistics.
We briefly recall it and then bring it in a form convenient for
deformation.

Let us consider a system with $n$ degrees of freedom
quantized in accordance with the parafermi or parabose statistics of
order $p$, i.e.,
a system of $n$ paraoscilators which is a
particular representation $\pi_{p}$ (of order $p$)
of the parastatistics algebra with trilinear exchange relations
(\ref{1}).

The Green ansatz states that the  parafermi (parabose) oscillators
$a^{+}_i$ and $a^{-}_{i}$  can be represented  as sums of $p$ fermi
(bose) oscillators
\beq
\pi_{p}(a^{\pm}_i)
=  \mathop{\sum}\limits_{r=1}^{p} a^{\pm (r)}_i \qquad \qquad
\label{Gran}
\eeq
  satisfying quadratic  commutation relations of the same type
(i.e., fermi for parafermi and bose for parabose)  for equal
indices   $(r)$
\beq
    [ a^{-(r)}_{i},\,a^{+(r)}_k ]_{\pm}=\delta_{ik},
    \qquad\quad [a^{-(r)}_{i}, a^{-(r)}_{k}]_{\pm}= [a^{+(r)}_{i},
a^{+(r)}_{k}]_{\pm}=0,
\label{ga1}
    \eeq
and  of the opposite type for the different indices
\beq
[a^{-(r)}_{i}, a^{-(s)}_{k}]_{\mp}= [a^{+(r)}_{i}, a^{+(s)}_k]_{\mp}=
[a^{-(r)}_{i},\,a^{+(s)}_{ k }]_{\mp}=0 \qquad r \neq s.
\label{ga2}
    \eeq
The upper (lower) signs stay  for the parafermi (parabose) case.

The coproduct endows the tensor product of $\cala$-modules of
the Hopf algebra $\cala$ with the structure of an $\cala$-module.
Thus one can use the coproduct  for constructing a
representation out of simple ones.
The simplest representations of
the parastatistics algebras are  the oscillator
representations  $\pi$  (with $p=1$).
Higher representations $\pi_p$
of parastatistics of order $p\geq 2$ arise through the
iterated coproduct
\cite{Pal3}.

Let us denote the ($p$-fold)  iteration of the coproduct
by ${}^{(}$\footnote{${}^{)}$The definition of $\Delta^{(p)}$  extended with the
counit $\epsilon$ is consistent with $\pi_0=\epsilon$}
\beq
\Delta^{(0)} = \epsilon, \quad \Delta^{(1)}= id, \quad
\Delta^{(2)} = \Delta, \quad  \ldots \quad \Delta^{(p)} =
( \underbrace{\Delta \otimes 1 \otimes \ldots \otimes 1}_{p-1})
\circ \Delta^{(p-1)}
\label{pfold}
\eeq
and $\pi$ denotes the projection from the (deformed) parafermi and
parabose algebra
onto the (deformed) fermionic $\mathfrak{F}$ ($\mathfrak{F}_{q}$) and
bosonic
$\mathfrak{B}$ ($\mathfrak{B}_{q}$) Fock representation, respectively.

\begin{proposition}
The Green ansatz is equivalent to the commutativity of the following
diagrams
\beq
\ba{ccc}
\pf(n) &   \stackrel{\Delta^{(p)}}{\longrightarrow}   &
\pf(n)^{\otimes p} \\
        & \pi_{p}  \searrow \qquad    &  \downarrow \pi^{\otimes p} \\
        &&  \mathfrak{F}(n)^{\otimes p}
\ea
\qquad \qquad
\ba{ccc}
\pb(n) &   \stackrel{\Delta^{(p)}}{\longrightarrow}   &
\pb(n)^{\otimes p} \\
       & \pi_{p}  \searrow \qquad   &  \downarrow \pi^{\otimes p} \\
        &&  \mathfrak{B}(n)^{\otimes p}
\ea
\label{dia}
\eeq
\end{proposition}

{\it Proof:}
Using  the coproduct of the Theorem \ref{Hopfpf} for $q=1$ and projecting
on the Fock representation we can choose the components of the Green
ansatz
to be the summands in the expressions
\beq
{\pi^{\otimes p} \circ \Delta^{(p)}}(a^{\pm}_i) =
\mathop{\sum}\limits_{r=1}^{p} \underbrace{1\otimes
\ldots \otimes  1}_{r-1  } \otimes \pi(a^{\pm}_i) \otimes
\underbrace{ 1 \otimes \ldots \otimes 1}_{p-r} :=
\mathop{\sum}\limits_{r=1}^{p}
a^{\pm (r) }_{ i }
\eeq
The check that the Green components $a^{\pm (r) }_{i\,}$
satisfy the bilinear commutation relations (\ref{ga1}) and (\ref{ga2})
is direct, however one has to keep in mind that the tensor product is
${\mathbb Z}_{2}$-graded in the parabose case and non-graded in the
parafermi
case, which explains why the anomalous commutation relations
(\ref{ga2})
appear. We emphasize that the grading of the tensor product turns out
to be the opposite to the (independent) grading
of the bose or fermi algebra which appears on each
site $(r)$.

The diagrams (\ref{dia}) are commutative if and only if
\beq
\pi_{p}(a^{\pm}_i) =  {\pi^{\otimes p} \circ \Delta^{(p)}} (a^{\pm}_i)
\eeq
which is exactly the statement of the Green ansatz (\ref{Gran}).
$\Box$

We are now in a position to extend the Green ansatz
to the  deformed parafermi $\pfq(n)$ and parabose $\pbq(n)$ algebras.
The simplest representation of  $\pfq(n)$ and  $\pbq(n)$ of
parastatistics order $p=1$,
are the  deformed fermionic $\mathfrak{F}_{q}$ and bosonic
$\mathfrak{B}_{q}$
Fock representations, respectively and let $\pi$ be the projection on
these Fock spaces.

  \begin{definition}
      The system of quadratic exchange relations stemming from the
      commutativity of the diagrams
\beq
\ba{ccc}
\pfq(n) &   \stackrel{\Delta^{(p)}}{\longrightarrow}   &
\pfq(n)^{\otimes p} \\
        & \pi_{p}  \searrow \qquad    &  \downarrow \pi^{\otimes p} \\
        &&  \mathfrak{F}_{q}(n)^{\otimes p}
\ea
\qquad \qquad
\ba{ccc}
\pbq(n) &   \stackrel{\Delta^{(p)}}{\longrightarrow}   &
\pbq(n)^{\otimes p} \\
        & \pi_{p}  \searrow \qquad   &  \downarrow \pi^{\otimes p} \\
        &&  \mathfrak{B}_{q}(n)^{\otimes p}
\ea
\label{qdia}
\eeq
is the   deformed Green ansatz of parastatistics of
order $p$.
Here $\Delta^{(p)}$ stays for the $p$-fold non-cocommutative
coproduct (\ref{pfold})  on the Hopf algebras
$\pfq(n)$ and $\pbq(n)$ 
  (see Theorem \ref{Hopfpf}).
\end{definition}

Let us show the consistency of the condition (\ref{pq}) with the
deformed Green
ansatz. The vacuum state $\rvac^{(p)}$ of the representation
$\pi_{p}$ is to
be identified with the tensor power of the oscillator ($p=1$) vacuum,
$\rvac^{(p)}= \rvac^{\otimes p}$.
Evaluating the iterated graded commutator (\ref{fe})
\beq
\Delta^{(p)}[\![a^{+}_i,\, a^{-}_{i} ]\!]= 
[\![\Delta^{(p)}a^{+}_i, \, \Delta^{(p)}a^{-}_{i} ]\!]=
\frac{(q^{h_{i}})^{\otimes p}  - (q^{- h_{i}})^{\otimes p}}
  {{q^{\demi}-q^{-\demi}}}
\eeq
on the vacuum state $\rvac^{\otimes p}$  in
the oscillator representations $\pi^{\otimes p}$
we get  the defining  condition (\ref{pq}) of the deformed $\pi_{p}$
\[
\mp \pi^{\otimes p} \circ \Delta^{(p)}[\![a^{+}_i,\, a^{-}_{i} ]\!]
\rvac^{(p)} =
\pi_{p}(a^{-}_{i}) \pi_{p}( a^{+}_i)  \rvac^{(p)}
=[p] \rvac^{(p)} \quad
(=\frac{q^{\frac{p}{2}}-q^{-\frac{p}{2}}}
{q^{\demi}-q^{-\demi}}
\rvac^{(p)} )
\]
since $\pi(q^{h_{i}})= q^{N_{i}\mp \demi}$, which proves the
consistency.

  The Green components   ${a}^{\pm (r) }_{i}$ 
  in a $\pfq(n)$ or $\pbq(n)$ representation $\pi_{p}$ of
parastatistics of order $p$
    will be chosen to be
  \beq
  \ba{rcl}
  a^{+(r)}_{i}&=& \pi^{\otimes p} \circ \Delta^{(r-1)} \otimes 1
\otimes
  \Delta^{(p-r)}
  \left( \sum_{k=1}^{n}L^{(+)}_{ik}\otimes  a^{+}_k \otimes 1\right)
  \\[5pt]
  a^{-(r)}_{i}&=&  \pi^{\otimes p} \circ \Delta^{(r-1)} \otimes 1
\otimes
  \Delta^{(p-r)}
  \left( \sum_{k=1}^{n} 1\otimes a^{-}_{k}\otimes L^{(-)}_{ki}\right)
  \ea
  \eeq
Note that the conjugation $\ast$ acts as reflection on the Green
indices $(r)$ 
$$(a^{\pm(r)}_i)^{\ast}= a^{\mp (r^{\ast})}_{i } \qquad \qquad
r^{\ast}=p-r+1.$$
        More explicitly
        the Green components look like
\beq
\ba{lclc}
a^{+(r)}_{i}&=&\mathop{\sum}\limits_{k_{1}, \ldots , k_{r}}
&{ \underline L}^{(+)}_{i k_{1}}\otimes { \underline L}^{(+)}_{ k_{1}
k_{2}}\otimes
\ldots \otimes { \underline L}^{(+)}_{ k_{r-1}  k_{r}}\otimes
{\underline a}^{+}_{k_{r}}
\otimes 1 
\ldots  \otimes 1 \\[6pt]
a^{-(r)}_{j }&=& \mathop{\sum}\limits_{  k_{1}, \ldots, k_{p-r}}&
1 \otimes 
\ldots \otimes 1 \otimes  {\underline a}^{-}_{k_{1}} \otimes
{ \underline L}^{(-)}_{k_{1} k_{2}}\otimes
{ \underline L}^{  (-)}_{k_{2} k_{3}}\otimes
\ldots \otimes
{ \underline L}^{(-)}_{ k_{p-r} j}
\label{gc}
\ea
\eeq
where the upper (lower) triangularity of the matrices
$L^{(+)}$($L^{(-)}$)
infers that only the terms subject to  the inequalities
$i \leq k_{1}\leq \ldots \leq k_{r}\leq  n $ are non-zero
( respectively ${n \geq k_{1}\geq \ldots \geq  k_{p-r} \geq j}$ ).
Unlike the non-deformed case each Green component ${a}^{\pm(r)}_{i}$
in the deformed
Green ansatz is a sum of many terms resulting from the mapping
${\pi^{\otimes p} \circ \Delta^{(p)}}$.

To present the results in a more concise form we
introduce the
      operators
         \beqa
     Q^{+(r)}_{ji} =& \pi^{\otimes p} \circ \Delta^{(r)}\otimes
\Delta^{(p-r)}  \,
     (\mathop{\sum}\limits_{k=1}^{n} L^{(+)}_{jk}
     \otimes L^{(-)}_{ki}) &\\
    {Q}^{-(r)}_{ji} =&\pi^{\otimes p} \circ \Delta^{(r-1)}\otimes
\Delta^{(p-r+1)} \,
      (\mathop{\sum}\limits_{k=1}^{n} L^{(+)}_{jk}
     \otimes L^{(-)}_{ki})&
     \label{cofor}
     \eeqa
    One readily sees that $(Q^{+(r)}_{ji})^{\ast}=
{Q}^{-(r^{\ast})}_{ij} $ and
    $({Q}^{-(r)}_{ji})^{\ast}=  {Q}^{+(r^{\ast})}_{ij}$.\\

We now summarize the deformed quadratic algebra anomalous
commutation rules.
  For different Green indices the Green components (\ref{gc}) quommute
  ( $[x , y  ]_{\pm q}= xy \pm q yx$) as  follows  (we suppose $r>s$)
      \beq
     \ba{lcrlccc}
     [ a^{+ (r)}_{i},a^{+ (s)}_{j} ]_{\mp}&= & \mp
     (q-q^{-1}) a^{+(r) }_{j}a^{+(s)}_{i}, &
       [ a^{-(r)}_{i},a^{-(s)}_{j}  ]_{\mp } &=& 0
        &\quad  i< j \\[4pt]
        [ a^{-(r)}_{i},a^{-(s) }_{j}]_{\mp}  &= &
     \pm (q-q^{-1}) a^{- (r)}_{j}a^{- (s)}_{i },
      &
      [ a^{+(r) }_{i},a^{+(s)}_{j}
     ]_{\mp}&=&0 &\quad  i> j
     \ea
\label{gr1}
     \eeq
\beq
     [ a^{+ (r)}_{i},a^{+ (s)}_{i}
     ]_{\mp q^{}} = 0
     \qquad   \qquad
     [ a^{-(r)}_{i},a^{-(s)}_{i}
     ]_{\mp q^{- 1}}=0 \qquad \qquad
      [ a^{-(r)}_{i },a^{+(s) }_{j}]_{\mp} = 0
     \eeq
     When the Green indices coincide one gets
		 \beq
      \ba{lclclcl}
     [a^{+ (r)}_{i}, a^{+(r) }_{j}]_{\pm q^{\mp \epsilon_{ij}}}&=&0
      & \quad &
      [a^{-(r)}_{i }, a^{-(r)}_{j }]_{\pm q^{\mp \epsilon_{ij}}}&=&0
\\[4pt]
      [a^{-(r)}_{i}, a^{+(r) }_{j}]_{\pm q^{ \mp 1}} &=&  q^{\mp
\frac{1}{2}}
     Q^{-(r)}_{ji} &\quad &
     [a^{-(r)}_{i}, a^{+(r)}_{j}]_{\pm q^{\pm 1}} &=&  q^{ \pm
\frac{1}{2}}
     {Q}^{+(r)}_{ji}
      \ea
     \eeq
where
    the operators $ Q^{+(r)}_{ji}$  and ${Q}^{-(r)}_{ji}$
   (\ref{cofor})
are quadratic in the Green components
\beq
\ba{crll}
q^{\mp \demi}{ Q}_{ji}^{-(r)}=&
(q-q^{-1}) \sum_{s={1}}^{r-1} q^{\mp (r-s)} a^{+(s) }_{j} a_{i}^{-(s)}
=& (q^{ \pm \demi}{Q}_{ij}^{+(r^{\ast})})^{\ast} &\quad
i>j \\[8pt]
q^{\mp \demi}{ Q}_{ji}^{-(r)}=&
-(q-q^{-1}) \sum_{s={r}}^{p} q^{\mp (r-s)} a^{+(s) }_{j} a_{i}^{-(s)}
=&(q^{\pm \demi}{Q}_{ij}^{+(r^{\ast})})^{\ast} &\quad
i<j
\ea
\eeq
\beq
\ba{lccr}
q^{\pm \frac{1}{2}} Q^{+(r)}_{ii}
= q^{\mp (r-\frac{p}{2} -\demi)} (q^{N_{i}})^{\otimes p } &-&
(q-q^{-1})
\sum_{s=r+1}^{p} &
q^{\mp (r-s)}  a^{+ (s)}_{i} a_{i}^{-(s)} \\[8pt]
q^{\mp \frac{1}{2}}{Q}^{-(r)}_{ii}
= q^{\mp (r-\frac{p}{2} -\demi)} (q^{-N_{i}})^{\otimes p }&+&
(q-q^{-1})
\sum_{s=1}^{r-1} &
q^{\mp (r-s)}  a^{+(s)}_{i} a_{i}^{-(s)}
\ea
\label{theend}
\eeq

  The system of relations (\ref{gr1}-\ref{theend}) with the upper (lower) signs
defines the
generalization of the Green ansatz for
         the deformed parafermi $\pfq(n)$  (parabose
$\pbq(n)$) algebras.

\section{Acknowledgements}
It is a pleasure to thank Michel Dubois-Violette,
Ludmil Hadjiivanov, Tchavdar Palev and Ivan Todorov for
 inspiring discussions and their interest in this work.
B.A. is very grateful to Julius Wess for the kind invitation to visit
the LMU-Munich, to Dieter L\"ust for the warm hospitality at the
Theory Division there and acknowledges the support of DFG (Deutsche
Forschungsgemeinschaft). T.P.  extends his gratitude to G\'erard
Duchamps and Christophe Tollu
for inviting him at LIPN in the team Combinatorics, Informatics $\&$
Physics.
T.P. was partially  supported by the  Bulgarian National Council
for Scientific Research under the project PH-1406 and  Euclid Network
HPRN-CT-2002-00325.

           \appendix
\section{Appendix}

		 {\it Proof of Lemma 1:}
All lowering $U_q(gl_n)$ Chevalley  generators $E_{-i}$ kill the
state $\Lambda^{n-1,n}_n$
$$
ad_{E_{-i}}\Lambda^{n-1,n}_n = 0 \qquad i=1,\dots , n-1.
$$
The states $\Lambda^{i,n}_j$ and $\tilde{\Lambda}^{i,j}_{n}$ for all admissible $i,j$  (\ref{hr})
arise
through the adjoint action of the raising $U_q(gl_n)$  generators
as seen from the diagram    in
         which the decorated arrows
denote the adjoint actions  $ad_{E_i}$
$$
\xymatrix{
\Lambda^{n-1,n}_{n}  \ar[d]^{E_{n-2}} \ar[r]^{E_{n-1}}&
               \tilde{\Lambda}^{n-1,n-1}_{n}\ar[r]^{E_{n-2}}&
 	\tilde{\Lambda}^{n-2,n-1}_{n}\ar[r]^{E_{n-3}} \ar[d]^{E_{n-2}}&
\ldots \ar[r]^{E_2}&\tilde{\Lambda}^{2,n-1}_{n} \ar[d]^{E_{n-2}} \ar[r]^{E_1}  &
\tilde{\Lambda}^{1,n-1}_{n} \ar[d]^{E_{n-2}}\\
	{\Lambda}^{n-2,n}_{n}  \ar[d]^{E_{n-3}} \ar[r]^{E_{n-1}}   &
 	{\Lambda}^{n-2,n}_{n-1}  \ar[d]^{E_{n-3}}\ar[r]^{E_{n-2}} &
	\tilde{\Lambda}^{n-2,n-2}_{n} \ar[r]^{E_{n-3}} &
 \ldots \ar[r]^{E_2}&\tilde{\Lambda}^{2,n-2}_{n}
\ar[d]^{E_{n-3}} \ar[r]^{E_1}  &
\tilde{\Lambda}^{1,n-2}_{n} \ar[d]^{E_{n-3}}\\
	 {\Lambda}^{n-3,n}_{n} \ar[d]^{E_{n-4}} \ar[r]^{E_{n-1}} &
{\Lambda}^{n-3,n}_{n-1} \ar[d]^{E_{n-4}} \ar[r]^{E_{n-2}}  &
 {\Lambda}^{n-3,n}_{n-2} \ar[d]^{E_{n-4}} \ar[r]^{E_{n-3}} &
 \ldots \ar[r]^{E_2}&\tilde{\Lambda}^{2,n-3}_{n}\ar[d]^{E_{n-4}} \ar[r]^{E_1}  &
\tilde{\Lambda}^{1,n-3}_{n} \ar[d]^{E_{n-4}}\\
\vdots \ar[d]^{E_1}	&
\vdots\ar[d]^{E_1} &
\vdots\ar[d]^{E_1}
&
&
\vdots &\vdots \ar[d]^{E_1}  \\
	 {\Lambda}^{1,n}_{n} \ar[r]^{E_{n-1}}&
 {\Lambda}^{1,n}_{n-1} \ar[r]^{E_{n-2}}&
 {\Lambda}^{1,n}_{n-2}\ar[r]^{E_{n-3}} &
  \ldots \ar[r]^{E_2}&
 {\Lambda}^{1,n}_{2} \ar[r]^{E_{1}}&\tilde{\Lambda}^{1,1}_{n}   }
	$$

 Next, the new state
        $\Lambda^{n-2,n-1}_{n-1}=ad_{E_{n-1}} {\Lambda}^{n-2,n}_{n-1}$
        stays at the top of a new diagram 
			of the same type	with  $n'=n-1$.
           By induction we obtain  all the states in $\call$ (\ref{hr}).
The state $\tilde{\Lambda}^{1,1}_2$ is the  highest weight of $\call$.
One can check that the adjoint $U_q(gl_n)$-action  does not
bring out of  $\call$ which completes the proof.

The $U_q(gl_n)$-module $\cal L$ is a smooth
deformation   of a
Schur module associated with the Young diagram $\lambda =(2,1)$\cite{thesis}. The  states  $\Lambda^{i,k}_j$ and $\tilde{\Lambda}^{i,j}_{k}$
in $\call$
are labelled with semistandard Young tableaux.
Hence the dimension is $\dim \call = \frac{(n+1)n(n-1)}{3}$.


\begin{thebibliography}{33}
 \bibitem{W} E. Wigner. {\em Phys. Rev. \/ \bf 77}(1950), 711-712.

\bibitem{Green} H. S. Green.
         {\em Phys. Rev. \/ \bf 90}(1953), 270-273.


 \bibitem{QHE}  B.I.Halperin, {\it Phys.Rev.Lett.}{\bf 52}(1984),
1583; \\
R.B.Laughlin, {\it Phys.Rev.Lett.}{\bf 60}(1988), 2677.


        \bibitem{anyons}
J.M.Leinaas, J.Myrheim, {\it Nuovo Cimento B} {\bf 37}(1977), 1; \\
F.Wilczek, {\it Phys.Rev.Lett.}{\bf 49}(1982), 957


\bibitem{quons}
Chi-Keung Chow, O.W.Greenberg, {\it Phys.Lett A}{\bf 283}(2001), 20;
\\
O.W.Greenberg, J.D.Delgado, {\it Phys.Lett. A}{\bf 288}(2001), 139.

\bibitem {Hadji} L. K. Hadjiivanov.
             {\em J. Math. Phys. \/ \bf 34  }(1993), 5476-5492.

\bibitem{Pal2} T. D. Palev.
              {\em Lett. Math. Phys. \/ \bf 31}(1994), 151-158.

\bibitem{Pal1} T. D. Palev.
{\em J.Phys.A:Math. Gen.\/ \bf 26}(1993), L1111-L1116.


\bibitem{Pal3} T. D. Palev. 
                { \em J.Phys. A \/ \bf27}(1994), 7373-7386.

\bibitem{Quesne}  C. Daskaloyannis, K. Kanakoglou, I. Tsohanatjis.
{\em J. Math. Phys. \/ \bf 41}(2000), 652. \\
C. Quesne.
               { \em Phys. Lett. A \/ \bf 260 }(1999), 437-440.

\bibitem{KT} S. Kamefuchi, Y. Takahashi.
              {\em Nucl. Phys. \/ \bf 36}(1962), 177-206. \\
C. Ryan, E.C.G. Sudarshan.
              {\em Nucl. Phys. \/ \bf 47}(1963), 206-211.
\bibitem{GP} A. Ganchev, T. Palev.
{\em J. Math. Phys. \/ \bf 21}(1980) 797.



\bibitem{Drin} V. G. Drinfeld.
         { \em Proceedings of the International Congress
        of Mathematicians, Berkley, 1986, A.M.Gleason (ed.)}
              (American Mathematical Society, Providence, RI, 1987)
              pp. 798-820.

\bibitem{Jimbo} M. Jimbo.
  {\em  Lett. Math. Phys. \/ \bf 10}(1985), 63-69.

\bibitem{FRT} L. Faddeev, N. Reshetikhin, L. Takhtajan.
         {\em Algebra i Analiz \/ \bf 1}(1989), 178-206,
        english translation: {\em Leningrad Math. J. \/ \bf 1}
  (1990), 193.

\bibitem{KhTo} S. M. Khoroshkin, V. N. Tolstoy.
              {\em Commun. Math. Phys. \/ \bf 141}(1991), 599-617.

\bibitem{Pal4} T. D. Palev, N.I. Stoilova.
                                                        { \em
Lett.Math.Phys. \/ \bf 28 }(1993), 187-194.

 \bibitem{Kac} V. Kac.  {\em Adv. Math. \/ \bf 26}(1977), 8.

 \bibitem{isa} A. P. Isaev. {\em J. Phys. A \/ \bf 29} (1996), 6903-6910.              


\bibitem{GM}   O. W. Greenberg, A.M.L.Messiah.
        {\em Phys.Rev. B \/ \bf 138}(1965), 1155-1167.

\bibitem{OK} Y. Ohnuki, S. Kamefuchi.
        Quantum field theory and parastatistics.
        Springer-Verlag 1982.

\bibitem {unita}  H.-D. Doebner, T.D. Palev, N.I. Stoilova
{\em J.Phys. A \/  \bf  35 }(2002) 9367 
;
\\
L. K. Hadjiivanov, R. R. Paunov, I.T. Todorov.
              {\em J. Math. Phys. \/ \bf 33}(1992), 1379.

\bibitem{thesis} T. Popov Ph.D. thesis (2003), LPT, Orsay and INRNE,
Sofia.  ({\em LPT preprint server})




\end{thebibliography}
                         \end{document}